\documentclass[12pt,preprint]{aastex}

\newcommand{\dpp}[2]{\frac{\partial #1}{\partial #2}}
\newcommand{\ddt}[1]{\dpp{#1}{t}}

\newcommand{\mi}[1]{\mbox{\boldmath$#1$}}
\newcommand{\Nbv}{\mathcal{N}^{\,2}}

\shorttitle{MHD waves in inclined magnetic field}
\shortauthors{Parchevsky et al.}

\begin{document}
\title{Numerical simulation of excitation and propagation of helioseismic
MHD waves in magnetostatic models of sunspots}

\author{K. Parchevsky, A. Kosovichev\altaffilmark{1},
E. Khomenko\altaffilmark{2,3}, V. Olshevsky\altaffilmark{3}, M.
Collados\altaffilmark{2}}

\affil{Stanford University, HEPL, Stanford CA 94305, USA}
\email{kparchevsky@solar.stanford.edu}

\altaffiltext{1}{Stanford University, HEPL, Stanford CA 94305, USA.}
\altaffiltext{2}{Instituto de Astrofisica de Canarias, 38205,
C/V\'ia L\'actea, s/n, Tenerife, Spain.} \altaffiltext{3}{Main
Astronomical Observatory, NAS, 03680 Kyiv, Ukraine.}

\begin{abstract}
We present comparison of numerical simulations of propagation of MHD
waves,excited by subphotospheric perturbations, in two different
("deep" and "shallow") magnetostatic models of the sunspots. The
"deep" sunspot model distorts both the shape of the wavefront and
its amplitude stronger than the "shallow" model. For both sunspot
models, the surface gravity waves ($f$-mode) are affected by the
sunspots stronger than the acoustic $p$-modes. The wave amplitude
inside the sunspot depends on the photospheric strength of the
magnetic field and the distance of the source from the sunspot axis.
For the source located at 9~Mm from the center of the sunspot, the
wave amplitude increases when the wavefront passes through the
central part of the sunspot. For the source distance of 12~Mm, the
wave amplitude inside the sunspot is always smaller than outside.
For the same source distance from the sunspot center but for the
models with different strength of the magnetic field, the wave
amplitude inside the sunspot increases with the strength of the
magnetic field. The simulations show that unlike the case of the
uniform inclined background magnetic field, the $p$- and $f$-mode
waves are not spatially separated inside the sunspot where the
magnetic field is strongly non-uniform. These properties have to be
taken into account for interpretation of observations of MHD waves
traveling through sunspot regions.
\end{abstract}


\section{Introduction}
Details of MHD wave propagation inside magnetized regions are very
important for understanding of interaction, scattering, and
conversion of seismic waves by sunspots in the Sun. The structure of
sunspots themselves is not well understood. \citet{Pizzo1986}
proposed a self consistent magnetostatic sunspot model. Later
\citet{Low1975} presented a self-similar model. These two models
(model of Pizzo at the top and model of Low at the bottom) were
joined by \citet{Khomenko2008}. It is well understood now that the
sunspot is dynamic, so several attempts have been made to build
numerically a stable sunspot model with the background flows
\citep{Hurlburt2000,Botha2008}. Recently \citet{Rempel2009} obtained
realistically looking penumbral structures with outflows.

Local helioseismology provides a tool for reconstruction of the
internal structure (both profiles of wave speed and flows) of
sunspots from Doppler observations. The magnetic field of sunspots
affect the result of the helioseismic inversion. For understanding
the helioseismic effects of the magnetic field it is important to
perform direct simulations of MHD waves in different models of
sunspots. The simulations are also used for producing artificial
data for calibration and testing of helioseismic measurements and
inversion algorithms. There is wide gallery of numerical simulations
of propagation of MHD waves inside sunspots illustrating various
observed phenomena: power deficit of $p$-modes, acoustic halos,
azimuthal phase shift variations of the signal in the sunspot
penumbra, and so on. Not all of these phenomena are caused by the
direct effects of magnetic fields. Indirect effects of magnetic
fields must be also taken into account. For instance,
\citet{Parchevsky2007b} showed that 50\% of the observed acoustic
power deficit in sunspots can be explained by the absence of the
acoustic sources inside sunspots, where strong magnetic field
inhibits convective motions, which are the primary source of solar
oscillations. The suppression of acoustic sources is one of the most
important indirect effects of the solar magnetoseismology. Another
example of the indirect effects are changes in the density and
temperature stratifications caused by magnetic fields. Among the
direct effects of magnetic field (effects caused by magnetic
stresses on wave perturbations) an important role is played by
conversion into different types of MHD waves. For instance, 2D
simulations of wave propagation in inclined magnetic fields (e.g
\citet{Cally1997,Cally2000,Spruit1992}) showed that fast MHD waves
can be converted into slow MHD waves, which can leave the
computational domain, and thus interpreted as an "absorption" of the
fast mode. Acoustic halos around sunspots were obtained in
simulations by \citet{Hanasoge2008} as a result of wave
transformations. However, \citet{Jacoutot2008} noted that this
effect can be explained by changes in the excitation properties of
solar convection in moderate field strength regions.
\citet{Cameron2008} obtained the peak restrictions of 3 kG for the
photospheric magnetic field strength from simulations of scattering
$f$-modes by the sunspot. Three dimensional simulations of linear
MHD wave propagation performed by \citet{Parchevsky2009} show that
the inclined magnetic field can be partly responsible for the
azimuthal variations of travel times around sunspots found by
\citet{Zhao2006} from helioseismic observations. Refraction of
upward propagating MHD waves in the solar atmosphere was simulated
by e.g. \citet{Khomenko2006,Cally2008}. Recently,
\citet{Khomenko2009} carried out 2D simulations of the interaction
of MHD waves with magnetostatic sunspot models. In this paper we
present initial results of our 3D modeling of this problem.

Results of linear numerical simulations of MHD waves propagation in
sunspots mainly depend on the setup of the problem: 2D or 3D
simulations, choice of the background model of the sunspot, the way
of excitation of the waves, and treating of the boundary conditions.
The goal of this paper is to study and compare properties of 3D
propagation of MHD waves in different magnetostatic models of
sunspots, and also describe how different types of waves are
affected by the sunspot. Waves are generated by localized
subphotospheric (depth of 100 km) sources of the vertical force. For
such sources simulated waves propagate along the same ray paths as
in the Sun. The numerical method and the background sunspot models
are described in \S2 and \S3. The results of simulations are
discussed in \S4.

\section{Governing equations}

Propagation of MHD waves inside the Sun is described by the
following system of linearized equations:
\begin{equation}\label{Eq:MHD_3D}
\begin{array}{l}
\displaystyle \ddt{\rho'} + \nabla\cdot \mi{m}'=0,\\[12pt]
\displaystyle \ddt{\mi{m}'} + \nabla p' - \frac{1}{4\pi}\left[
(\nabla\times\mi{B}')\times\mi{B}_0 +
(\nabla\times\mi{B}_0)\times\mi{B}'\right] = \rho'\mi{g}_0 + \mi{S}(r,t),\\[12pt]
\displaystyle
\ddt{\mi{B}'}=\nabla\times\left(\frac{\mi{m}'}{\rho_0}\times\mi{B}_0\right),\\[12pt]
\displaystyle \ddt{p'} + c_{s0}^2\nabla\cdot \mi{m}' +
c_{s0}^2\frac{\Nbv_0}{g_0}m_z = 0,
\end{array}
\end{equation}
where $\mi{m}'=\rho_0\mi{v}'$ is the momentum perturbation,
$\mi{v}'$, $\rho'$, $p'$, and $\mi{B}'$ are the velocity, density,
pressure, and magnetic field perturbations respectively,
$\mi{S}(r,t)=(0,0,S_z(r,t))^T$ is the wave source function. The
quantities with subscript 0, such as gravity $\mi{g}_0$, sound speed
$c_{s0}$, and Br\"unt-V\"ais\"al\"a frequency $\mathcal{N}_0$
correspond to the background model, $\mi{B}_0$ is the background
magnetic field satisfying the usual magnetohydrostatic equilibrium
equation. The spatial and temporal behavior of the wave source is
modeled by function $S_z(r,t)= A H(r)F(t)$:
\begin{eqnarray}
H(r) &=& \left\{
\begin{array}{ll}
\displaystyle
\left(1-\frac{r^2}{R_{src}^2}\right)^2 & \mbox{if } r\leq R_{src}\\
\displaystyle 0 & \mbox{if } r>R_{src}
\end{array}
\right.\label{Eq:SourceXYZ}\\
F(t)&=& \left(1-2\tau^2\right)e^{-\tau^2}.\label{Eq:SourceT}
\end{eqnarray}
where $R_{src}$ is the source radius,
$r=\sqrt{(x-x_{src})^2+(y-y_{src})^2+(z-z_{src})^2}$ is the distance
from the source center, $\tau$ is given by equation
\begin{equation}
\tau=\frac{\omega_0 (t-t_0)}{2} - \pi, \qquad t_0\leq t\leq
t_0+\frac{4\pi}{\omega},
\end{equation}
where $\omega_0$ is the central source frequency, $t_0$ is the
moment of the source initiation. This source model provides the wave
spectrum, which closely resembles the solar spectrum. It has a peak
near the central frequency $\omega_0$ and spreads over a broad
frequency interval. The source spectrum is:
\begin{equation}
|\hat{F}(\omega)| \equiv \left|\int_{-\infty}^\infty F(t)e^{-i\omega
t} dt\right|
=4\sqrt{\pi}\;\frac{\omega^2}{\omega_0^3}\;e^{-\frac{\omega^2}{\omega_0^2}}.
\end{equation}
A superposition of such sources with uniform distribution of central
frequencies randomly distributed below the photosphere describes
very well the observed solar oscillation spectrum
\citep{Parchevsky2007a}.

For numerical solution of Eqs (\ref{Eq:MHD_3D}) a semi-discrete
finite difference scheme of high order was used. At the top and
bottom boundaries non-reflective boundary conditions based on the
Perfectly Matched Layer (PML) technique were set. Details of
numerical realization of the code can be found in
\citet{Parchevsky2007a}.

\section{Background model of the sunspot}
We used two types of axially symmetric magnetohydrostatic background
models of the sunspot described in \citet{Khomenko2008} ("shallow"
model) and \citet{Khomenko2005} ("deep" model). The "shallow"
sunspot model is obtained as a combination of the self-similar
solution \citep{Low1975} in deep layers with the solution of Pizzo
\citep{Pizzo1986} in upper layers. We calculated three instances
(a), (b), and (c) of the "shallow" model with the following
photospheric strengths of the magnetic field at the sunspot axis:
0.83~kG, 1.4~kG, and 2.2~kG respectively. Strengths of the magnetic
field at the bottom of the domain for these models are 5.0~kG,
8.0~kG, and 12.5~kG respectively. The depth of the domain (from the
photosphere) for the "shallow" models is 9.87~Mm. The position of
the photosphere for the "quiet" Sun (at the boundary of the sunspot)
coincides with the photospheric level in the standard model S
\citep{Christensen-Dalsgaard1996}. For comparison purposes we
calculated one instance of the "deep" sunspot model with the maximum
strength of the magnetic field of 843~G at the photosphere and 29 kG
at the bottom of the domain. The "deep" sunspot model is based on
solution of Pizzo everywhere in the domain. The depths of the deep
model is 7.5~Mm.

\begin{figure}
\centering \includegraphics[height=20 cm]{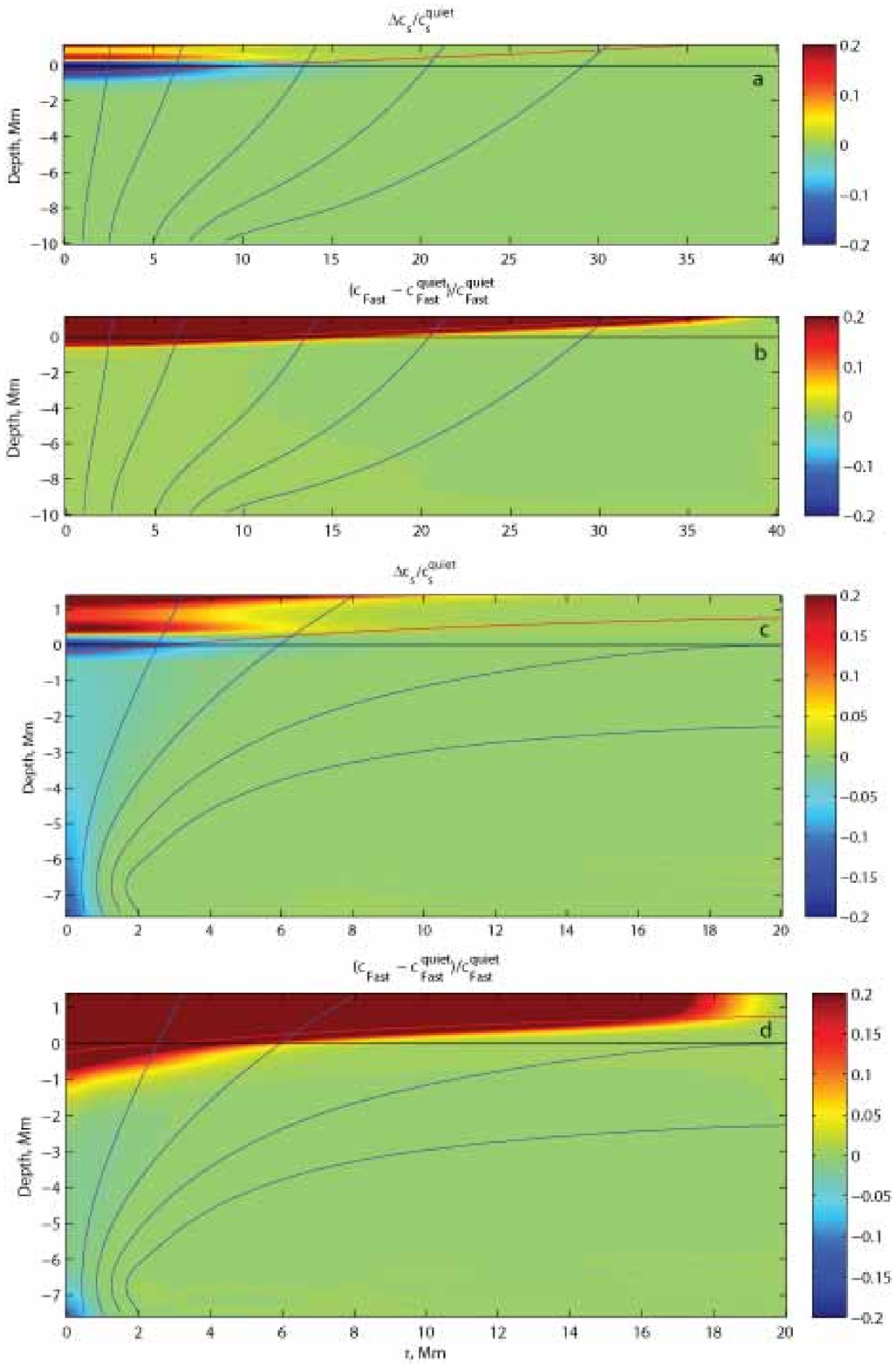}
\caption{Maps of the relative sound speed perturbation $\Delta c/c$
in the "shallow" (panel a) and the "deep" (panel c) sunspot models.
Panels (b) and (d) show maps of relative speed perturbations of fast
MHD waves for the "shallow" and "deep" models respectively. The
solid horizontal black line and the red curve represent the level of
the photosphere of the quiet Sun and level $\beta$~=~1
respectively.} \label{Fig:csMap}
\end{figure}

Maps of the relative sound speed perturbations for the "shallow"
(panel a) and "deep" (panel c) models with photospheric strength of
the magnetic field of 843 G and 836 G respectively are presented in
Figure \ref{Fig:csMap}. The black horizontal line marks the position
of the photosphere. The red curve shows position of $\beta$~=~1
level. Both types of the models were calculated under the assumption
that $\Gamma_1$~=~5/3. Panels b and d show variations of relative
speed of the fast MHD wave. The velocity of the fast MHD wave
depends on the angle between the wave vector and the vector of the
magnetic field. The maximum value is plotted. So, panels (a) and (b)
represent bottom and top limits (depending on the direction of
propagation) for the speed of the fast MHD wave for the "shallow"
model. Panels (c) and (d) represent speed limits for the "deep"
model.

\begin{figure}
\includegraphics[width = 0.47\textwidth]{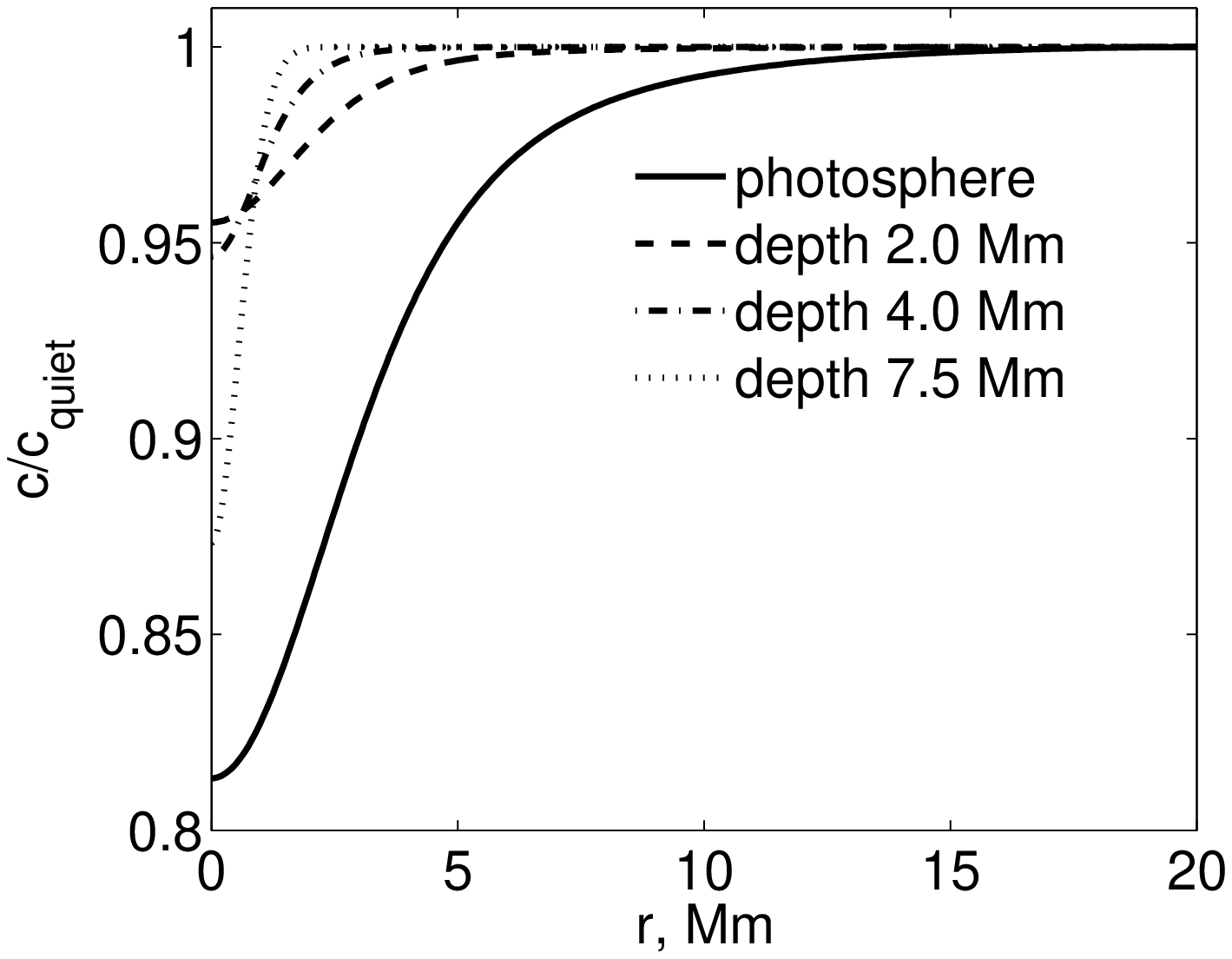} \hfill
\includegraphics[width = 0.47\textwidth]{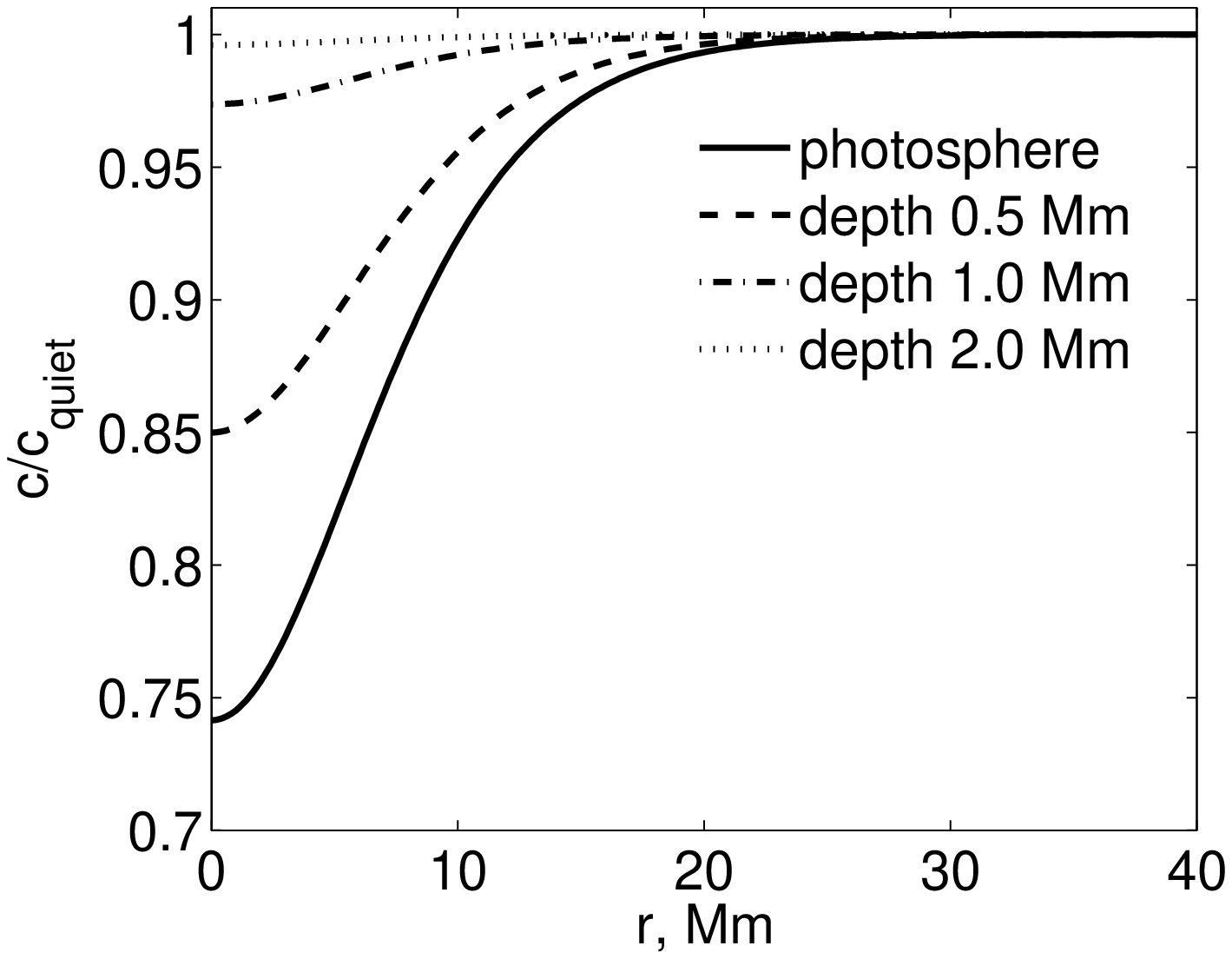}
\caption{Perturbations of the sound speed for different depth for
the "deep" model (left panel) and the "shallow" model (right panel).
For the "shallow" model horizontal variations of the sound speed is
negligible below 2~Mm.} \label{Fig:csHorz}
\end{figure}

\begin{figure}
\includegraphics[width = 0.47\textwidth]{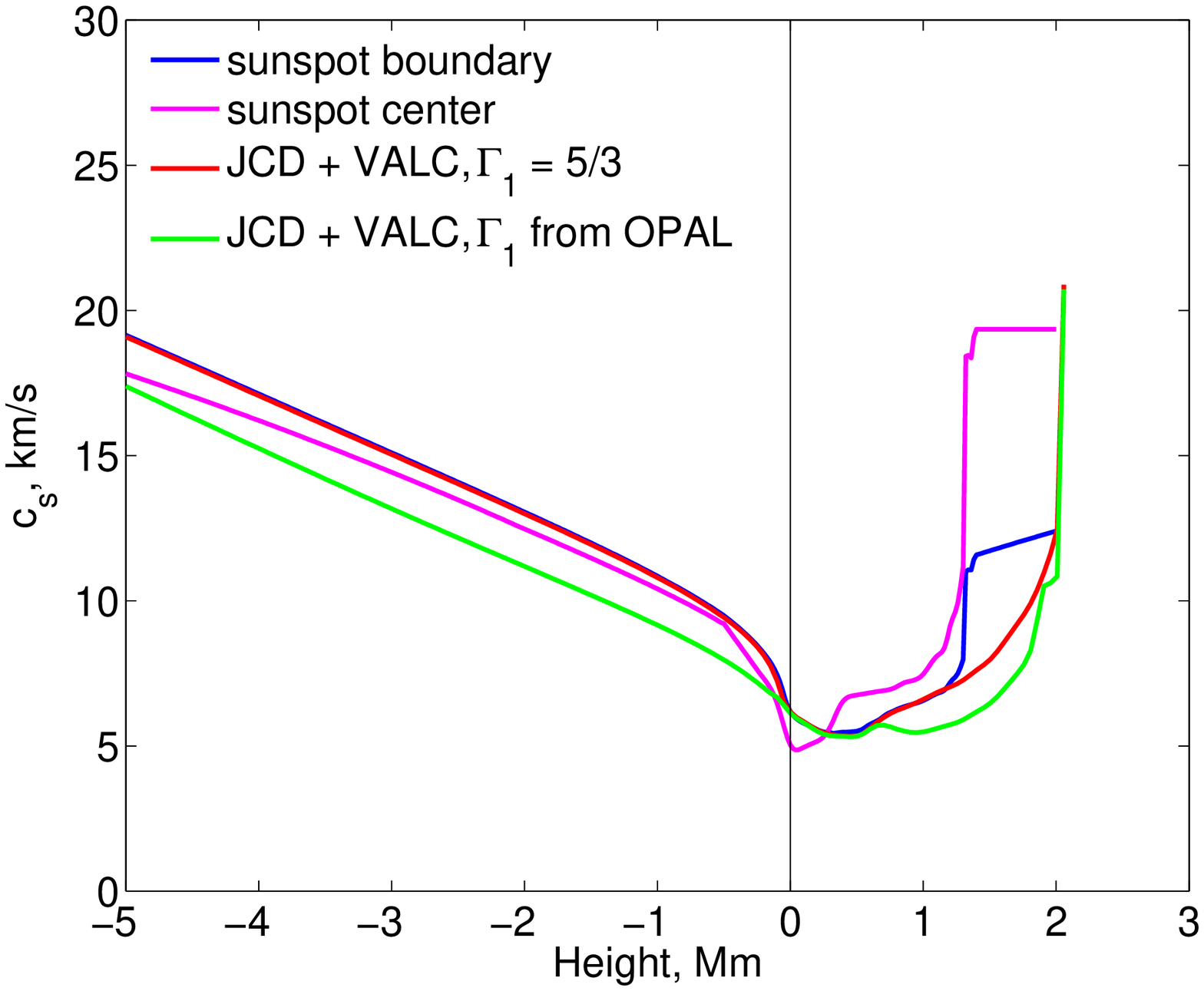} \hfill
\includegraphics[width = 0.47\textwidth]{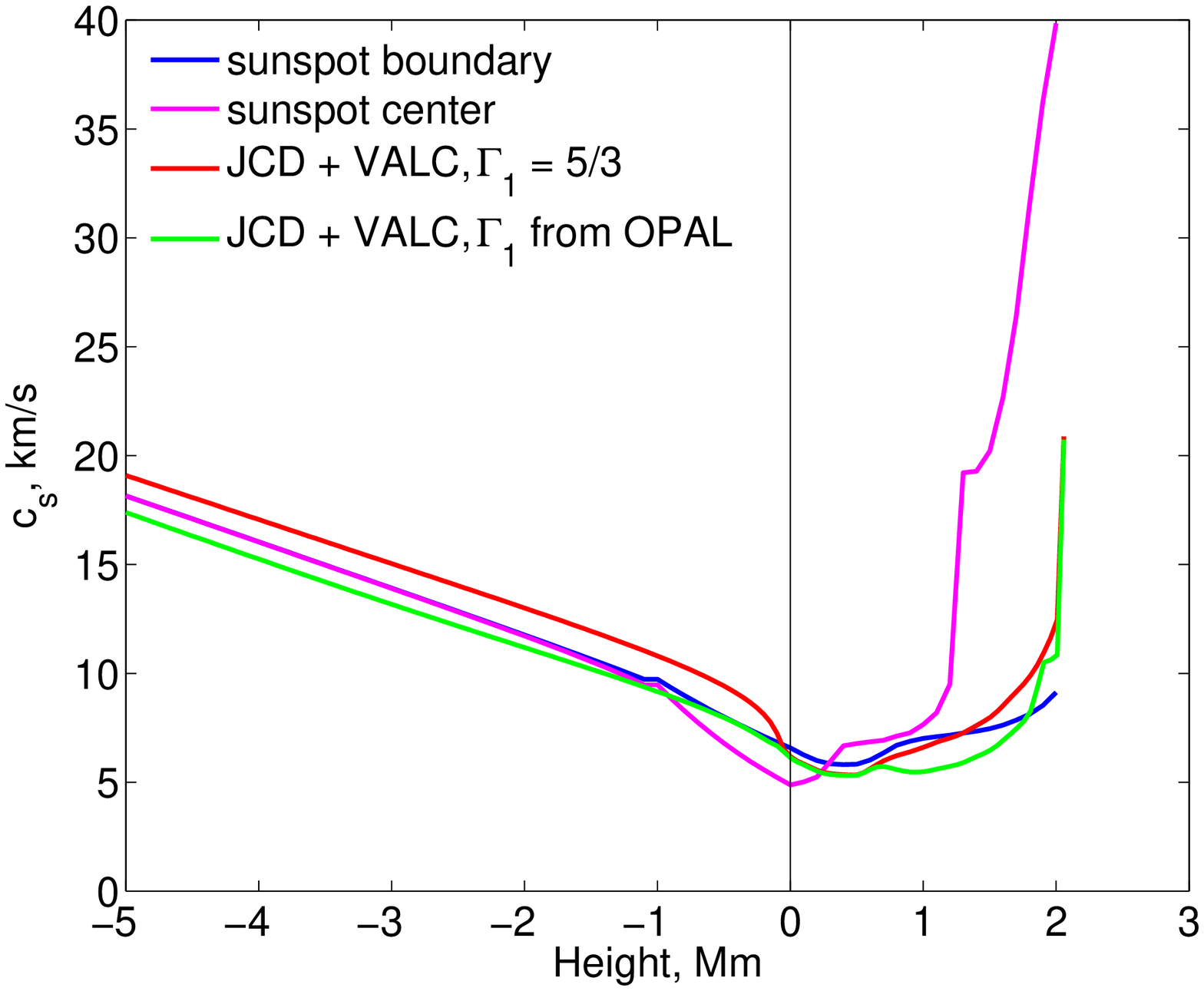}
\caption{Vertical profiles of perturbations of the sound speed for
the "deep" (left) and the "shallow" (right) sunspot models. The
magenta curve represents the vertical profile at the sunspot center.
The blue curve shows the sound speed profile at the sunspot
boundary. For comparison we plotted the sound speed profile from the
standard solar model S by Christensen-Dalsgaard with smoothly joined
VALC model of the chromosphere (green). Adiabatic exponent
$\Gamma_1$ is calculated from the OPAL equation of state. The same
curve, but with $\Gamma_1$ = 5/3 is shown by red color.}
\label{Fig:csVert}
\end{figure}

There are three main differences between these models: (i) the
topology of the magnetic field, (ii) the strength of the magnetic
field near the bottom of the domain, and (iii) dependence of the
horizontal profile of the sound speed on the depth. The magnetic
field lines (shown by blue curves in both panels of Figure
\ref{Fig:csMap}) are convex near the photosphere for the "deep"
model while in the "shallow" model the field lines are concave near
the photosphere. One dimensional cuts of the horizontal profiles of
the sound speed for different depths are shown in Figure
\ref{Fig:csHorz}. Profiles are scaled by the sound speed $c_{quiet}$
taken at the same depth of the outer boundary of the sunspot. In the
"shallow" model sound speed perturbation $\delta c/c =
c/c_{quiet}-1$ is close to zero everywhere starting from the depth
of about 2~Mm. In the "deep" model the sound speed perturbation is
non zero even at the bottom of the domain at 7.5~Mm. This means,
that in the "shallow" model waves, propagating at distances greater
than 10 Mm, propagate mostly in the region with the sound speed
profile of the quiet Sun below the region with the perturbed sound
speed. The vertical sound speed profiles at the sunspot center and
at the boundary for both models are shown in
Figure~\ref{Fig:csVert}.

The sound speed profiles along with the magnetic field structure are
the most significant parameters affecting propagation of the MHD
waves through the sunspot.

\section{Results and discussion}
In this section we present results of simulation of MHD waves
generated by a single source in different magnetohydrostatic
self-consistent background models. Axially symmetric models of
sunspots were interpolated on Cartesian grid with $\Delta x = \Delta
y = 0.15$~Mm. Vertical grid is non-uniform. We have three instances
(a), (b), and (c) of the "shallow" model (discussed above) with the
grid size of 376$\times$376$\times$67 and depth of 9.87~Mm (from the
photosphere). The vertical z-grid step varies from $\Delta z =
0.05$~Mm at the level of the photosphere and above  to $\Delta z =
0.52$~Mm at the bottom of the domain. Time step $\Delta t = 0.05$~s
is the same for simulations with all three instances of the
background "shallow" model. The "deep" model has the grid size of
184$\times$184$\times$62, depth of 7.5~Mm (from the photosphere),
and $\Delta t = 0.1$~s. The vertical z-grid step varies from $\Delta
z = 0.05$~Mm at the level of the photosphere and above to $\Delta z
= 0.4$~Mm at the bottom of the domain.

The source of the vertical component of force described above with
the spectrum given by Eq.~(5) is placed at the distance of 9~Mm from
the axis of the sunspot for the "shallow" models and 6~Mm for the
"deep" model. The depth of the source is 0.1~Mm for both models. The
top absorbing PML is placed at the height of 0.5~Mm and extends up
to 0.9~Mm. In this region the vertical profile of the sound speed at
the outer boundary of the sunspot is close to the profile in the
standard solar model with $\Gamma_1$~= 5/3. The lateral boundary
conditions are chosen to be periodic.

\begin{figure}
\includegraphics[width=1.0\textwidth]{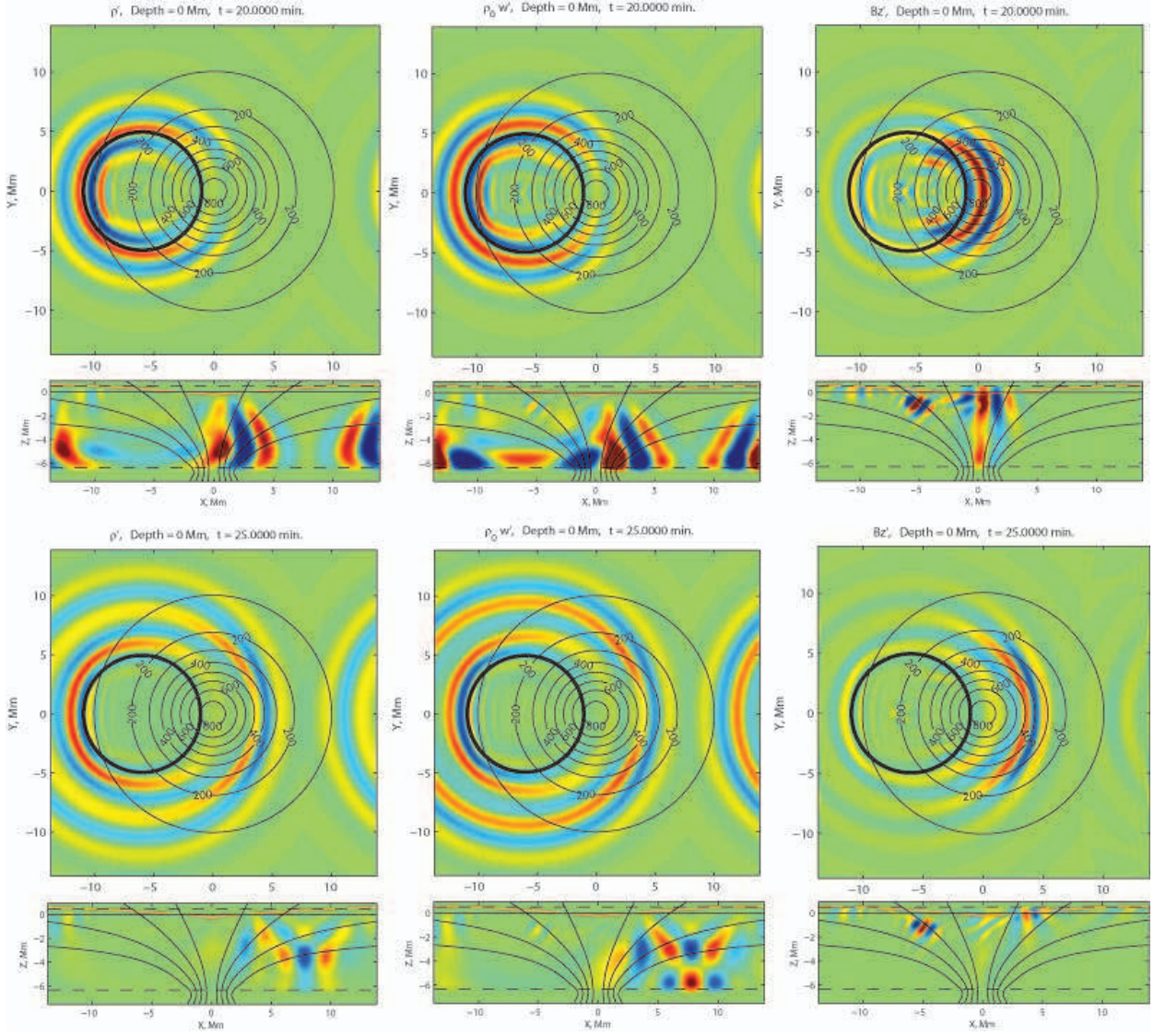}
{\caption{Snapshot of density $\rho'$ (left), z-momentum $\rho_0 w'$
(middle), and $B_z'$ (right) perturbations in the wave, propagating
in the "deep" model, at moments $t$=20 min. (top) and $t$ = 25 min.
(bottom). Each panel consists of two pictures: the horizontal slice
of the domain at the photospheric level (top) and t vertical cuts of
the domain (bottom). The solid black horizontal lines and red curves
in the vertical cuts represent the position of the quiet photosphere
and the level $\beta$~=~1 respectively. Solid black curves represent
the magnetic field lines.}\label{Fig:OldSpot_map}}
\end{figure}

Snapshots of the simulated wavefield for the "deep" model are shown
in Figure~\ref{Fig:OldSpot_map}. Panels a, b, and c represent maps
of perturbations of density $\rho'$, z-momentum $\rho_0 w'$, and
vertical component of the magnetic field $B_z'$ respectively at
moment $t$~= 20 min. Panels d, e, and f represent maps of the same
variables at $t$~= 25 min. Each panel consists of horizontal
XY-slice at the photospheric level (top) and vertical XZ-slice
through the center of the sunspot (bottom). The white line in the
XZ-slice shows position of the photosphere. Solid black curves
represent the magnetic field lines. The dashed lines near the top
and bottom of the XZ-slice show the position of the damping layers.
For the reference we plotted a circle with radius of 5 Mm and origin
at the wave source location. The non-uniform background model causes
anisotropy of the amplitude of the wave front. When the wave front
reaches the center of the sunspot the amplitude of density and
momentum perturbations decrease. After passing the center of the
sunspot the amplitude restores its original value. At some moments,
the amplitude inside the sunspot can become greater than the
amplitude of wave propagating outside the sunspot. The shape of the
wave front is changed as well. This is more noticeable for inner
parts of the wave front, located inside the circle in panels a, b,
and c. Later we show that these parts of the wave front are formed
mostly by $f$-modes. The shape of the outer parts of the wave front
(formed mostly by $p$-modes) remains close to the circle. A wave,
propagating along the magnetic field lines is appeared at the source
location. It is clearly seen in vertical slices of $\rho_0w'$ and
$B_z'$, but absent in $\rho'$. This wave consists of a mixture of
the Alfven and slow MHD waves.

\begin{figure}
\includegraphics[width=1.0\textwidth]{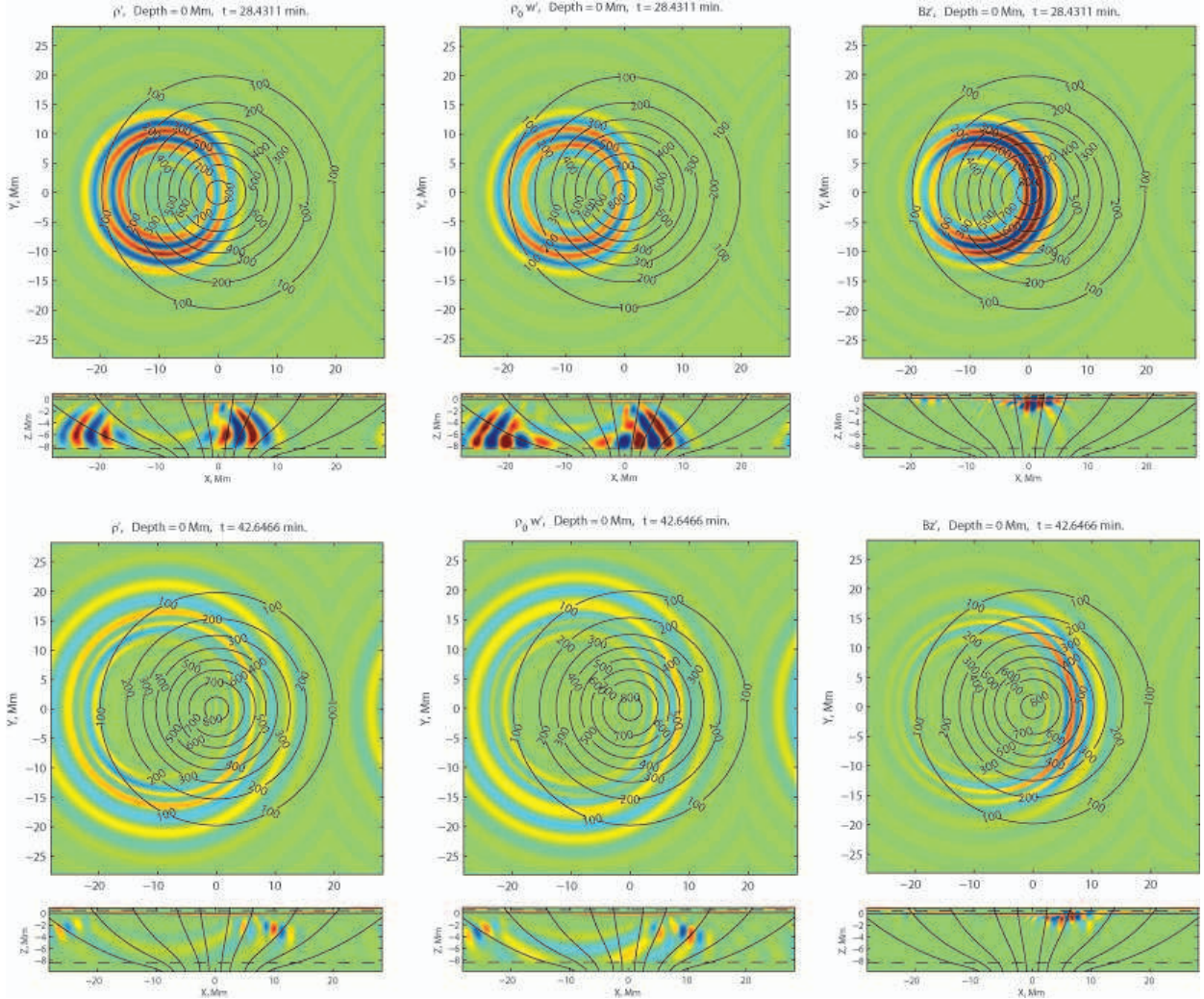}
{\caption{Snapshot of density $\rho'$ (left), z-momentum $\rho_0 w'$
(middle), and $B_z'$ (right) perturbations in the wave, propagating
in the "shallow" model with the photospheric strength of the
magnetic field of 0.83~kG, at moments $t$=28.4 min. (top) and $t$ =
42.6 min. (bottom). The solid black horizontal lines and the red
curves in the vertical cuts represent the position of the quiet
photosphere and the level $\beta$~=~1 respectively. Solid black
curves represent the magnetic field lines.}\label{Fig:NewSpot_map}}
\end{figure}

It is interesting to compare propagation of MHD waves in the "deep"
and "shallow" models with similar photospheric strength of the
magnetic field. Results of wave simulations in the "shallow" model
with the photospheric strength of the magnetic field $B_0$~= 0.83~kG
are shown in Figure~\ref{Fig:NewSpot_map}. There are common features
with simulations for the "deep" model and there are differences. In
general, in the "shallow" model waves show the same behavior as in
the "deep" model. The amplitude of  $\rho'$ and $\rho_0w'$ decreases
when the wave front reaches the center of the sunspot, but the ratio
of amplitudes is closer to unity than in the "deep" model. After
passing the center of the sunspot the wave restores its amplitude.
The amplitude of momentum perturbations $\rho_0w'$ becomes slightly
bigger inside the sunspot than outside, but again, not as much as in
the "deep" model. Perturbation of density $\rho'$ remains smaller
inside the sunspot than outside for all moments of time. The wave
that contains a mixture of the Alfven and slow MHD waves is
generated near the source location, as in the "deep" model, but the
amplitude of this wave is much smaller and it is not seen in
Figure~\ref{Fig:NewSpot_map}. The shape of the wave front is more
circular than in simulations for the "deep" model. In general, we
can say, that the "shallow" model affects the waves less than the
"deep" model.

\begin{figure}
\includegraphics[width=1.0\textwidth]{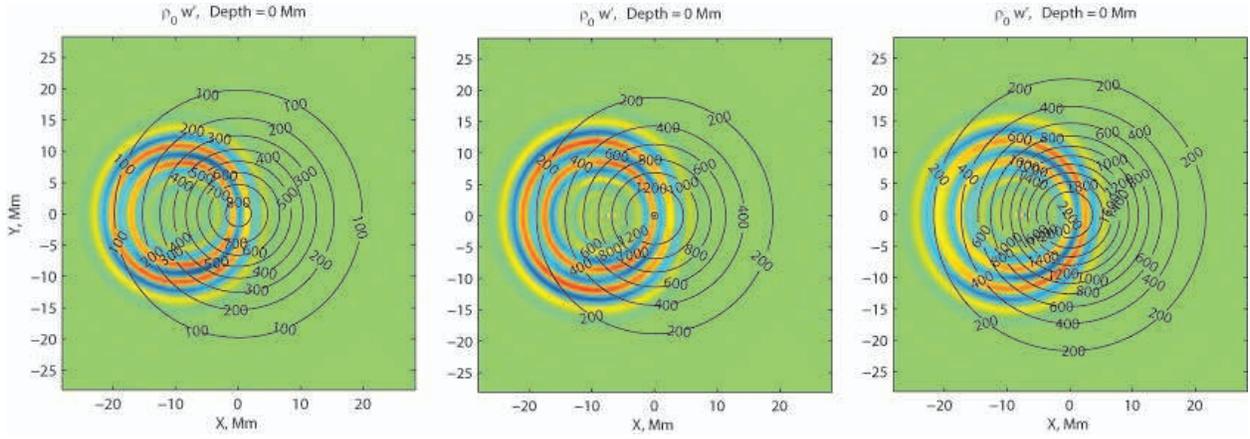}
{\caption{Dependence of z-momentum amplitude $\rho_0 w'$ on the
photospheric strength of the magnetic field $B_{ph}$ = 0.83~kG,
1.4~kG, 2.2~kG in the "shallow" models for panels a, b, and c
respectively. }\label{Fig:NewSpt_Bdep}}
\end{figure}

Comparison of the z-momentum maps at the same moments of time in
different instances of the "shallow" model with different strength
of the photospheric magnetic field is shown in
Figure~\ref{Fig:NewSpt_Bdep}. The wave amplitude inside the sunspot
grows with the strength of the magnetic field and for the surface
strength of 2.2~kG the wave amplitude inside the sunspot becomes
bigger than outside.

\begin{figure}
\includegraphics[width =
0.47\textwidth]{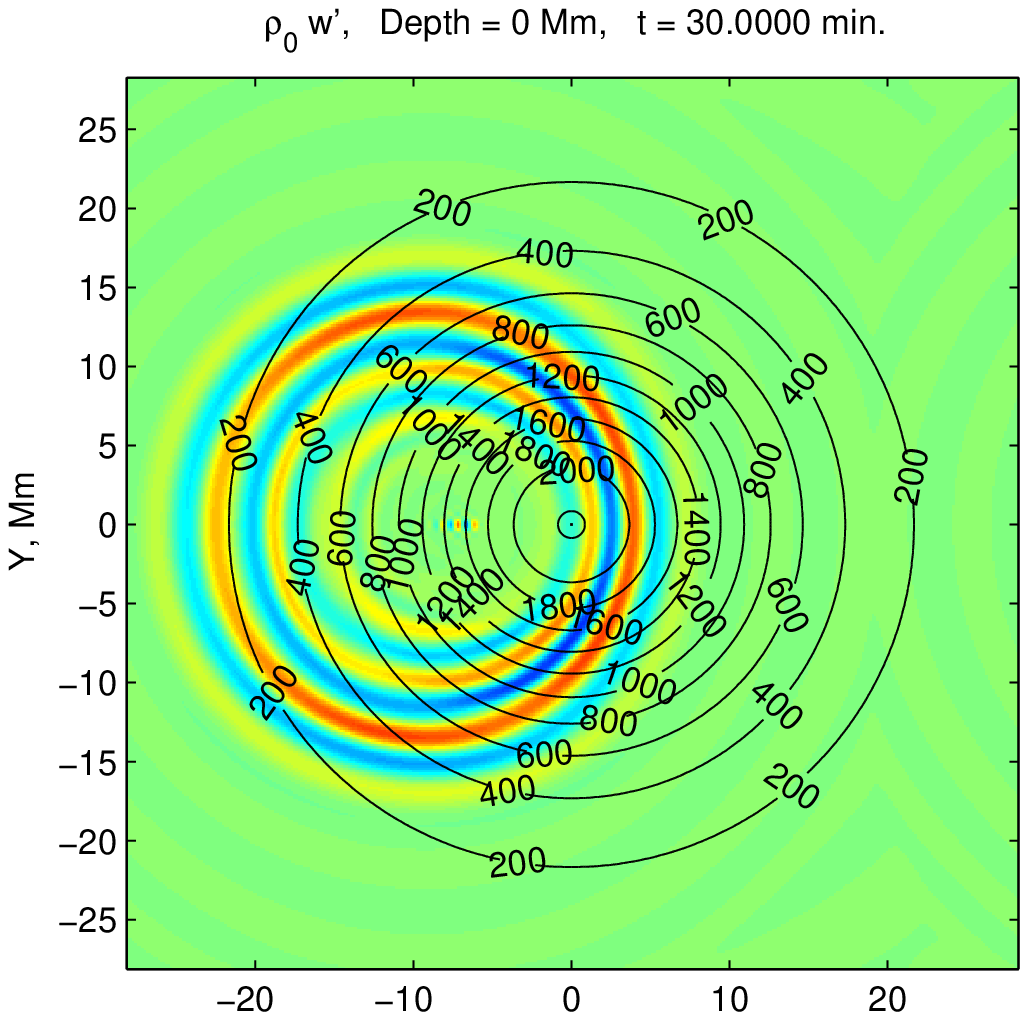}
\includegraphics[width = 0.47\textwidth]{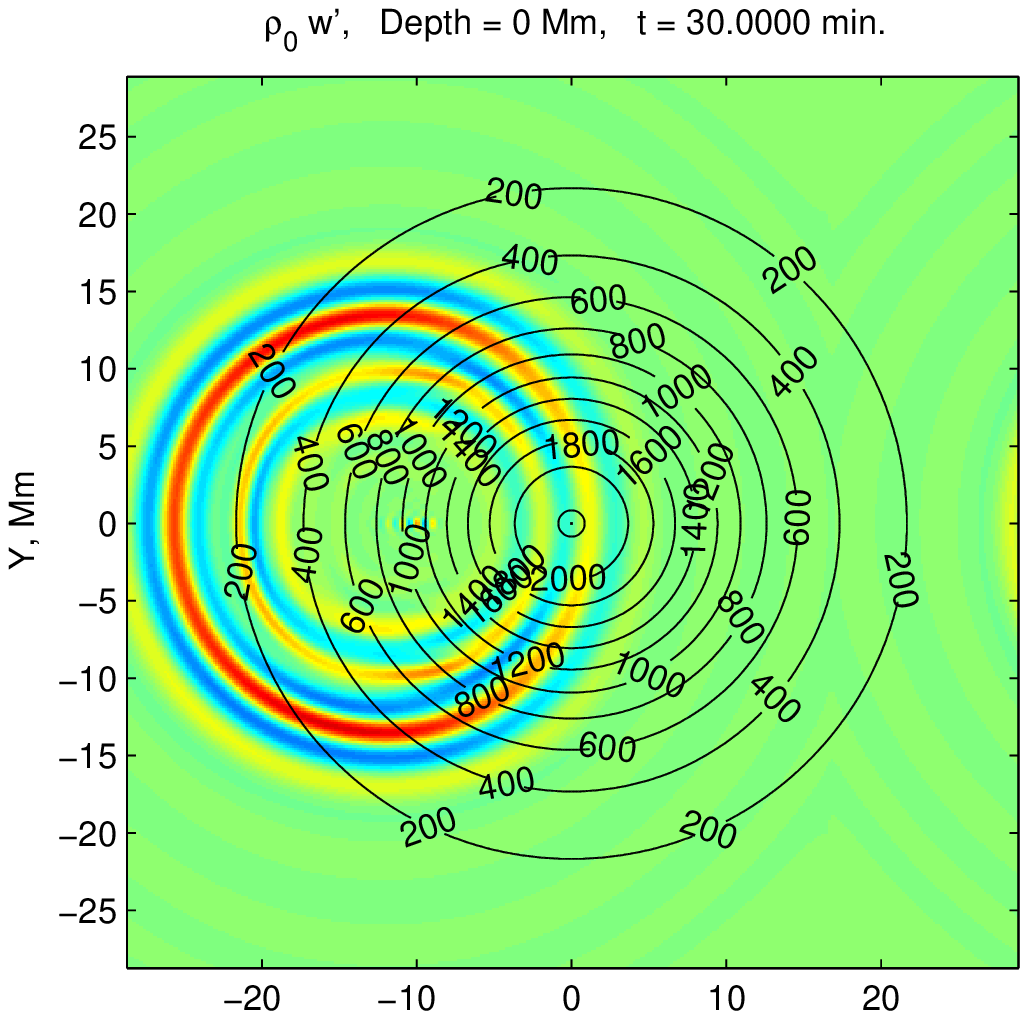}
\caption{Wave fields generated by sources located at 9~Mm (left
panel) and 12~Mm (right panel) from the sunspot axis in "shallow"
models respectively. The photospheric strength of the magnetic field
in both cases is 2.2~kG.} \label{Fig:NewSpt_9-12Mm}
\end{figure}

Behavior of MHD waves inside sunspots depends on the distance of the
wave source from the sunspot center. This happens because
magnetoacoustic waves in the Sun propagate through the solar
interior. This means that waves, generated by the source located
farther from the sunspot center, propagate deeper (in the region
with stronger magnetic field) than waves from the closer source. The
angle between the direction of wave propagation and the background
magnetic field is also different in these two cases. Snapshots of
the wave field for the "shallow" model with surface strength of the
magnetic field 2.2~kG are shown in Figure~\ref{Fig:NewSpt_9-12Mm}.
The left panel represents the wave generated by the source located
at 9~Mm from the sunspot axis. In the right panel the source is
located at 12~Mm from the sunspot center. The wave amplitude from
the close source at some moment of time becomes bigger inside the
sunspot than outside. The amplitude of the wave front of the distant
source remains smaller inside the sunspot than outside for all
moments of time.

\begin{figure}
\includegraphics{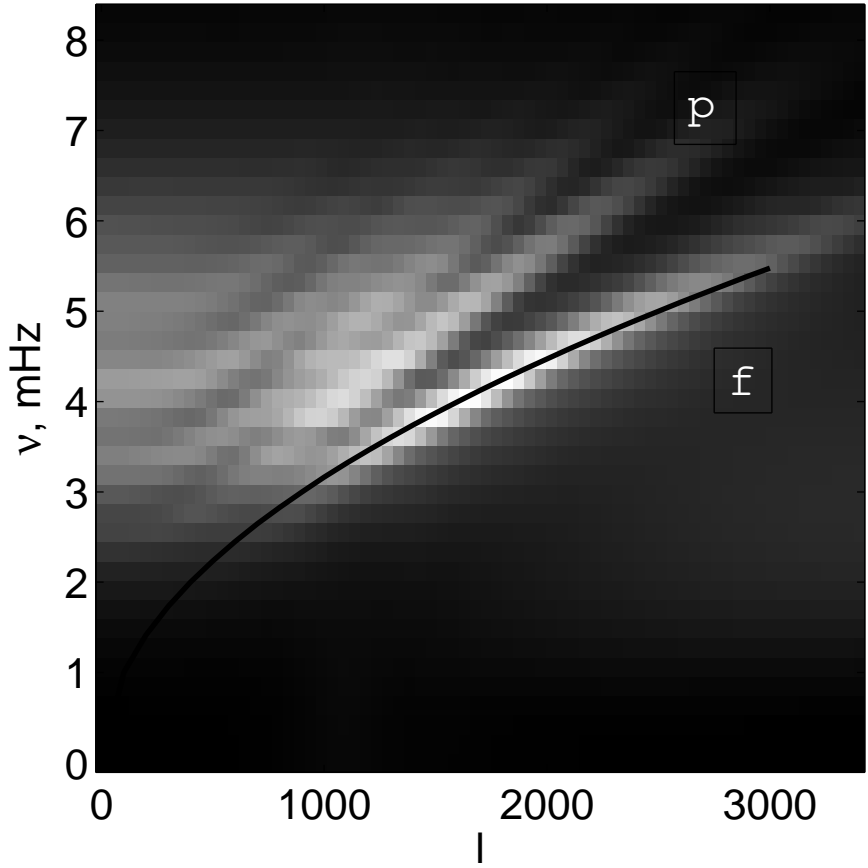} \caption{Spectrum
($k$-$\nu$ diagram) of the z-component of velocity for model~I. The
solid black curve shows the theoretical ridge of the $f$-mode in
absence of the magnetic field.} \label{Fig:kw_OldSpt}
\end{figure}

\begin{figure}
\centering\includegraphics[width=0.8\textwidth]{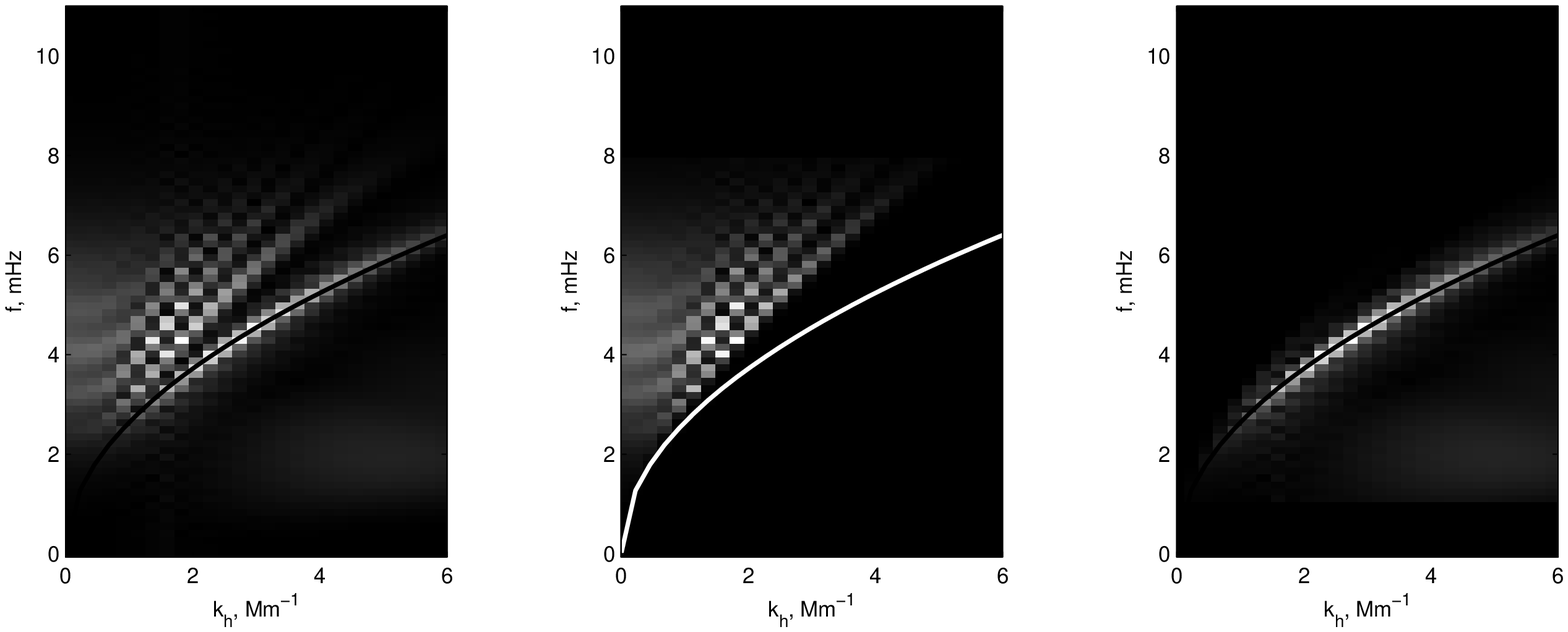}
\includegraphics[width=0.8\textwidth]{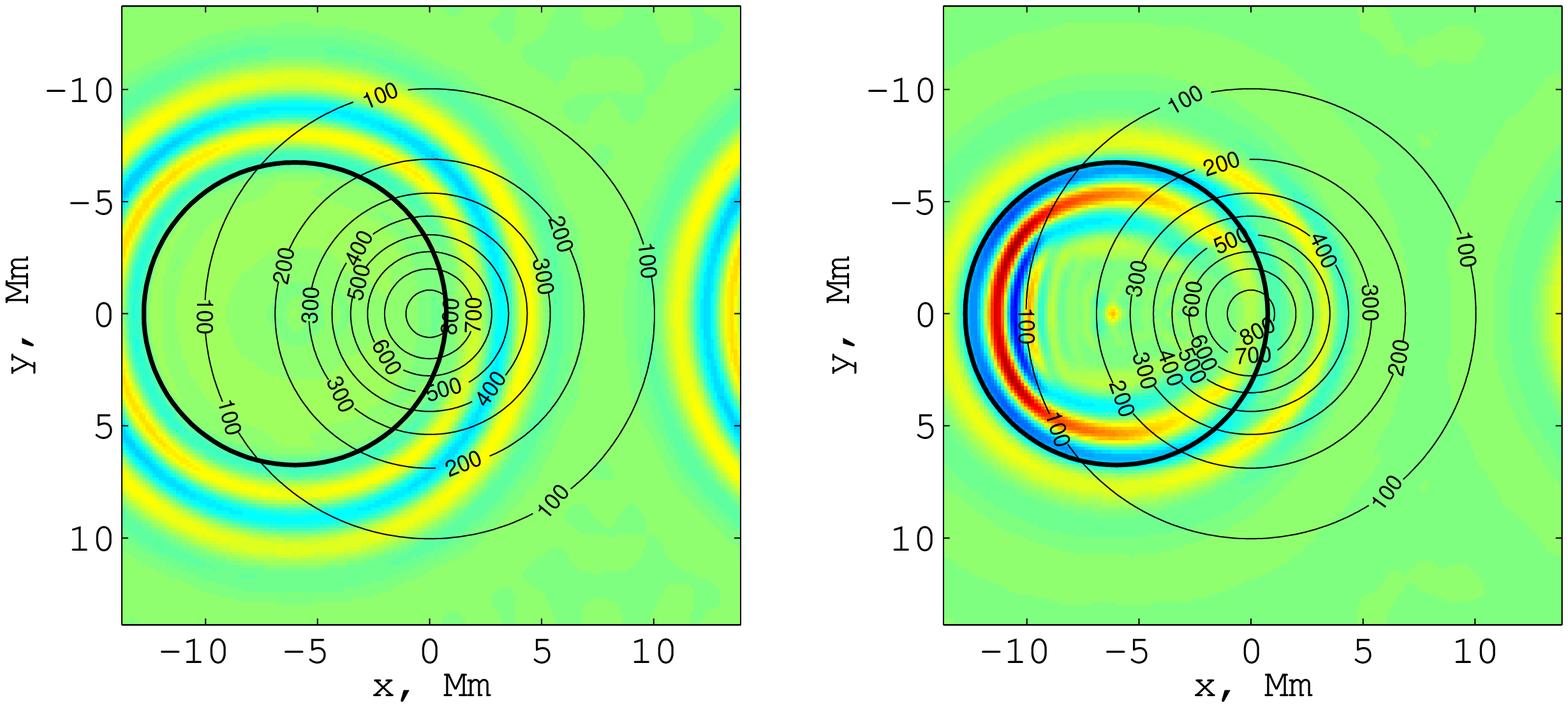}
\includegraphics[width=0.8\textwidth]{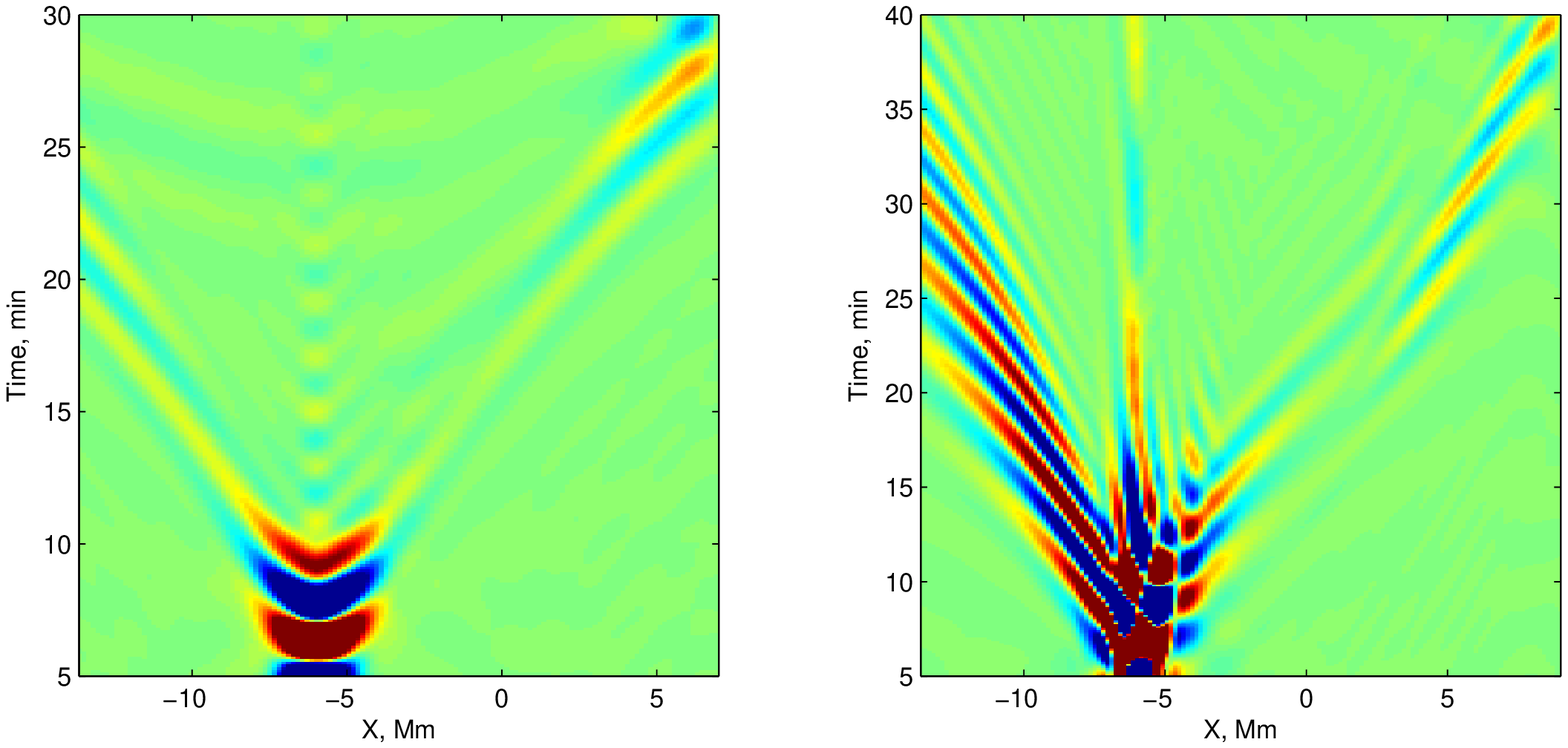}
\caption{Separation of $p$- and $f$-modes by filtering. The top row
represents the original and filtered $k$-$\nu$ diagrams. The middle
row represents the corresponding maps of z-component of velocity at
the moment of $t=$ 23 min. for $p$- and $f$-modes respectively. The
solid circle marks the inner part of the $p$-mode wavefront. The
bottom row shows time-distance diagrams for $p$- (left) and
$f$-modes (right).} \label{Fig:fp_kwseprt_OldSpt}
\end{figure}

The wave source generates a mixture of fast, slow, Alfven, and
magnetogravity waves. The spectrum ($k-\nu$ diagram) for this case
is shown in Figure~\ref{Fig:kw_OldSpt}. The solid black curve
represents the theoretical curve of the $f$-mode in absence of the
magnetic field. The fast MHD and magnetogravity modes can be
separated by filtering out the $f$-mode. Result of such separation
is shown in Figure~\ref{Fig:fp_kwseprt_OldSpt}. The top row
represents the original $k$-$\nu$ diagram (obtained form the
z-component of velocity at the photospheric level) and $k$-$\nu$
diagrams obtained by filtering out $f$- and $p$-modes respectively.
The white solid curve shows the theoretical position of the $f$-mode
ridge in the absence of the magnetic field. The middle row
represents maps of $V_z$ for $p$- and $f$-modes at the photospheric
level at the moment $t$=23 min. It was shown \citep{Parchevsky2009}
that in case of the horizontally uniform background model and
uniform inclined background magnetic field wavefronts of $p$- and
$f$-modes are spatially separated after 20-30~min. This does not
happen in the case of the non-uniform background magnitohydrostatic
model. The solid black circle marks the inner part of the wavefront
of $p$-modes. Although the amplitude of the $f$-modes (the right
panel) is mostly concentrated inside the circle, there is noticeable
non-zero amplitude in the region outside the circle where $p$-modes
present. It is clear, that the deformation of the wavefront (both
the shape and wave amplitude) due to the interaction with the
sunspot is much stronger for $f$-modes than for $p$-modes. The wave
amplitude of the $p$-mode wavefront decreases at the sunspot center
and quickly restores its value when the front passes through the
center. The amplitude of the $f$-mode wavefront remains perturbed at
all moments of time.

\section{Conclusion}
According to our 3D numerical simulations of MHD wave propagation in
two different models (referred as "shallow" and "deep") we can point
out to the following characteristic behavior of waves inside
sunspots. The interaction with the sunspot changes the shape of the
wave front and amplitude of the $f$-mode waves significantly
stronger than of the $p$-mode waves. The amplitude of the wave front
of the $p$-modes decreases when the wave reaches the sunspot center
and restores its original value when the wave passes the center of
the sunspot. The "shallow" model of sunspot affects waves less than
the "deep" model. This means that the horizontal inhomogeneity of
the sound speed profile inside the sunspot is mostly responsible for
perturbations of the wave front. In the "shallow" model the
horizontal distribution of the sound speed below 2 Mm is almost
uniform and the sound speed coincides with the value in the quiet
Sun at the same depth. Only magnetic field perturbs the wave front
in this region and the total perturbation of the wave front becomes
weaker than in the "deep" case.

Inside the sunspot magnetoacoustic and magnetogravity waves are not
spatially separated unlike the case of the uniform inclined magnetic
field. The wave amplitude inside sunspots depends on the strength of
the magnetic field and the distance of the wave source from  the
sunspot axis. The stronger photospheric magnetic field, the bigger
wave amplitude inside the sunspot (if the source is located at the
same distance from the sunspot center). For the source located at
9~Mm from the sunspot axis the wave amplitude inside the sunspot at
some moment becomes bigger than the amplitude outside. For the
source located at 12 Mm the wave amplitude inside the sunspot
remains smaller that outside for all moments of time.

In this paper, we presented initial results of 3D simulations of
helioseismic MHD waves in magnetostatic sunspot models.Future work
includes simulations with multiple random sources for testing the
travel-time measurement procedures of time-distance helioseismology,
and also modeling wave propagation in the MHD models of sunspots,
including flows \citep{Botha2008, Hurlburt2000}.


\end{document}